\documentclass[12pt]{article}
\usepackage{amsmath}
\usepackage[psamsfonts]{amssymb}
\usepackage[dvips]{graphicx}
\usepackage[hypertex]{hyperref}

\numberwithin{equation}{section}
\newcommand{\Zb}{\mathbb{Z}}

\newcommand{\Ncal}{\mathcal{N}}

\newcommand{\Hcal}{{\cal H}}

\DeclareMathOperator*{\Tr}{{\rm Tr}}

\DeclareMathOperator{\ch}{{\rm ch}}
\DeclareMathOperator{\chtilde}{\widetilde{\rm ch}}

\begin{document}
\thispagestyle{empty}

 \renewcommand{\thefootnote}{\fnsymbol{footnote}}
\begin{flushright}
 \begin{tabular}{l}
 {\tt arXiv:0809.0175[hep-th]}\\
 {SNUTP 08-007}
 \end{tabular}
\end{flushright}

 \vfill
 \begin{center}
 {\bfseries \Large Superconformal defects in the tricritical Ising model}
\vskip 1.9 truecm

\noindent{{\large
  Dongmin Gang \footnote{arima275(at)snu.ac.kr}
and Satoshi Yamaguchi \footnote{yamaguch(at)phya.snu.ac.kr} }}
\bigskip
 \vskip .9 truecm
\centerline{\it School of Physics and Astronomy,
Seoul National University,
Seoul 151-747, KOREA}
\vskip .4 truecm
\end{center}
 \vfill
\vskip 0.5 truecm

\begin{abstract}
We study superconformal defect lines in the tricritical Ising model in 2 dimensions.  By the folding trick, a superconformal defect is mapped to a superconformal boundary of the $\Ncal=1$ superconformal unitary minimal model of $c=7/5$ with $D_6-E_6$ modular invariant. It turns out that the complete set of the boundary states of $c=7/5$ $D_6-E_6$ model cannot be interpreted as the consistent set of superconformal defects in the tricritical Ising model since it does not contain the ``no defect'' boundary state.  Instead, we find a set of 18 consistent superconformal defects including ``no defect'' and satisfying the Cardy condition.
This set also includes some defects which are not purely transmissive or purely reflective.
\end{abstract}

\vfill
\vskip 0.5 truecm

\setcounter{footnote}{0}
\renewcommand{\thefootnote}{\arabic{footnote}}

\newpage

\tableofcontents
\section{Introduction and summary}

Conformal defects or interfaces in a conformal field theory are a kind of generalizations of conformal boundary conditions\cite{Wong:1994pa,Oshikawa:1996dj,Petkova:2000ip,Petkova:2001ag,Bachas:2001vj}. They describe the universality classes of the domain wall at which two different or same conformal field theories are connected. These conformal defects appear several different contexts in physics.  They describe impurities in condensed matter physics.  They also appear in the string theory; in AdS/CFT correspondence some branes in AdS spacetime correspond to defects in conformal field theory (see for example \cite{Kirsch:2004km}). Partial list of recent works on the conformal defects includes \cite{Frohlich:2006ch,Quella:2006de,Fuchs:2007tx,Brunner:2007qu,Bachas:2007td,Brunner:2007ur,Bajnok:2007jg,Brunner:2008fa,Runkel:2008gr}.

One of the main tools to investigate conformal defects is the folding trick\cite{Wong:1994pa,Oshikawa:1996dj}.
By this prescription, the problem is mapped into the conformal boundary problem in the direct product theory.

However it is not easy to get the classification of the boundary states in the folded theory even if the original theory is two minimal models.  This is because the product of two minimal models is not a minimal model in general, and the conformal boundary problem is not soluble in general.  The systematic ways to treat these boundary states are limited. One way is to consider the tensor products of the boundary states in each side.  These tensor product states are classified because the conformal boundaries in minimal models are classified\cite{Cardy:1989ir,Behrend:1999bn}.  These tensor product states are purely reflective defects in the unfolded theory.  The other way is to consider the permutation boundary states\cite{Recknagel:2002qq} when two CFTs in each sides are the same.  These states are purely transmissive defects (topological defects) in the unfolded theory.  Therefore it is not easy in general to obtain the defects which are not purely reflective or purely transmissive.

There are a few exceptions. For example conformal defects between two critical Ising models are mapped into conformal boundaries of certain $c=1$ CFT\cite{Oshikawa:1996dj}. This boundary problem is actually solved. There are a few more examples in which the conformal defects problem can be systematically treated\cite{Bachas:2001vj,Quella:2006de,Brunner:2007qu}.

In this paper we address defects between two tricritical Ising models\cite{Blume:1971,Friedan:1984rv}. The tricritical Ising model has $\Ncal=1$ superconformal symmetry. Actually it is the first model of the $\Ncal=1$ superconformal unitary minimal series ($m=3$ in eq.\eqref{1c}).

The direct product of two tricritical Ising models is not a minimal model, so it is difficult to classify all the conformal defects. However, when we require superconformal symmetry, the situation changes. The direct product of two tricritical Ising models with spin structure aligned is again a $\Ncal=1$ superconformal minimal model: $c=7/5$ ($m=10$ in eq.\eqref{1c}) $D_6-E_6$ modular invariant theory.  Thus we can treat this problem systematically.

At first sight, one seems to be able to solve the superconformal defect problem by just classifying the conformal boundary in this auxiliary theory using the Cardy condition\cite{Cardy:1989ir,Behrend:1999bn}. However it turns out not to work. There is no ``no defect'' boundary state in this classification which is expected to exist.

In this paper, we employ the following two as the criteria for the consistent set of superconformal defects.
\begin{enumerate}
 \item It includes ``no defect.''
 \item It satisfies the Cardy condition.
\end{enumerate}
As a result, we found 18 superconformal defects (see eq.\eqref{18defects}). This set of superconformal defects includes purely transmissive ones and purely reflective ones as well as intermediate ones. We calculated transmission coefficient (see eq.\eqref{transmission}), introduced by \cite{Quella:2006de}, for these defects.

The construction of this paper is as follows. In section \ref{sec-review} we review general techniques treating defects: the folding trick, the Cardy condition and so on.  Section \ref{sec-main} is the main section of this paper where we find 18 superconformal defects in the tricritical Ising model.  We collect some properties of $\Ncal=1$ superconformal minimal models in appendix \ref{app-minimal}. In appendix \ref{app-m=10} we classify all the boundary state in $c=7/5$ $D_6-E_6$ theory.

\section{(Super)conformal defects in 2-dimensional conformal field theories}
\label{sec-review}
In this section, we review some basic tools to treat the (super)conformal defects in two dimensional CFTs, like the folding trick and boundary states.

\subsection{(Super)conformal defects and the folding trick}
Consider a defect line on the real axis between two conformal field theories ( $\textmd{CFT}$s ): $\textmd{CFT}_1$ defined on upper half plane and $\textmd{CFT}_2$ on the lower plane.
The defect is called `conformal' if the current generating translation tangential to the defect is preserved across the defect.
And it is called `superconformal' if supercurrents $G,\bar{G}$ are preserved across the defect.  These conditions are written as
\begin{align}
T^{(1)}(z) - \bar{T}^{(1)}(\bar{z}) &= T^{(2)}(z) - \bar{T}^{(2)}(\bar{z}) |_{\textrm{at the defect}} , \nonumber
\\
G^{(1)}(z) - \eta \bar{G}^{(1)}(\bar{z}) &= \xi(G^{(2)}(z) - \eta\bar{G}^{(2)}(\bar{z})) |_{\textrm{at the defect}} ,
\label{defect-glue}
\end{align}
where $\eta,\xi=\pm 1$.

There are two extremal cases of the gluing conditions \eqref{defect-glue}. One is purely transmissive defects; holomorphic and anti-holomorphic currents are continuous across the defects individually. This kind of defects is sometimes called ``topological defects'' in the literature.
When two $\textmd{CFT}$s are the same, the simple example of purely transmissive defect  is ``no defect.'' The other is totally reflecting defects; each side of \eqref{defect-glue} is zero.
In this case, the two $\textmd{CFT}$s are decoupled and the defects can be considered as (super)conformal boundaries of each $\textmd{CFT}$.

In order to treat defects, it is convenient to use ``folding trick.''
By folding the two $\textmd{CFT}$s along the defects, we get the folded theory $\textmd{CFT}_1\otimes \overline{\textmd{CFT}}_2$ with boundary. $\overline{\textmd{CFT}}$ means the CFT obtained by exchanging holomorphic and anti-holomorphic degrees of freedom in $\textmd{CFT}$. The defect becomes the boundary of this folded theory. Actually the gluing conditions \eqref{defect-glue} can be rewritten as
\begin{align}
T^{(1)}(z) + \bar T^{(2)}(z^*) &= \bar{T}^{(1)}(\bar{z}) + T^{(2)}(\bar{z}^*)|_{\text{at the boundary}}, \nonumber
\\
G^{(1)}(z) + \xi\eta \bar G^{(2)}(z^*) &= \eta(\bar{G}^{(1)}(\bar{z}) + \xi\eta G^{(2)}(\bar{z}^*))|_{\textrm{at the boundary}}.
\end{align}
These conditions are the (super) conformal boundary condition that the boundary states in $\textmd{CFT}_1 \otimes \overline{\textmd{CFT}}_2$ should satisfy.
Therefore defects between two $\textmd{CFT}$s can be considered as boundary states in the folded theory.

In this paper we only treat left-right symmetric theory i.e. $\textmd{CFT}=\overline{\textmd{CFT}}$. So we just write $T^{(2)}(z)$ instead writing $\bar T^{(2)}(z^*)$.

\begin{figure}
\begin{center}
 \includegraphics[width=7truecm]{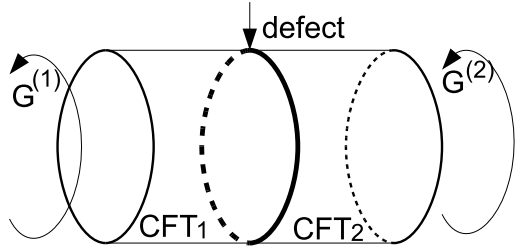}
\end{center}
\caption{Two half cylinders connected by the defect. The spin structure (periodicity of SUSY current $G$) must be the same in order to preserve the supersymmetry.}
\label{fig-spin}
\end{figure}

Here we make a remark about a subtlety for the ``direct product'' of two superconformal field theory.  {\em The naive direct product of two superconformal field theories is not a superconformal field theory.} This is because there is a sum of two SUSY currents with two different spin structures (NS or R) in the naive direct product theory.  Actually in order to satisfy eq.\eqref{defect-glue}, $G^{(1)}$ and $G^{(2)}$ (and $\bar G^{(1)}$ and $\bar G^{(2)}$) must have the same spin structures (see figure \ref{fig-spin}).

Therefore when considering superconformal defect, we will employ the direct product theory with aligned spin structure, denoted by $D(\textmd{CFT}_1 \otimes \textmd{CFT}_2)$, as the auxiliary theory instead of the naive direct product theory. This is justified as follows. The defect operator $F$ is a map from the Hilbert space of CFT$_1$ to that of CFT$_2$. Since $F$ preserve the supersymmetry, it must preserve the spin structure i.e. the periodicity of the SUSY current. Namely it maps an NSNS state to an NSNS state and an RR state to an RR state; it does not map an NSNS state to an RR state or an RR state to an NSNS state. Thus, $F$ can be written as
\begin{align}
 F=\sum_{a,b \in {\rm NSNS}}c_{a,b}|a\rangle \langle b|
+\sum_{a',b'\in {\rm RR}}c_{a',b'}|a'\rangle \langle b'|.
\end{align}
with some coefficients $c_{a,b},c_{a',b'}$.
After folding, $F$ becomes a state in the folded theory as
\begin{align}
  |F\rangle=\sum_{a,b \in {\rm NSNS}}c_{a,b}|a\rangle\otimes | b\rangle
+\sum_{a',b'\in {\rm RR}}c_{a',b'}|a'\rangle\otimes | b'\rangle.
\end{align}
This state $|F\rangle$ can be embedded in $D(\textmd{CFT}_1 \otimes \textmd{CFT}_2)$, because the spin structure is aligned in $|F\rangle$. The gluing condition \eqref{defect-glue} implies that $|F\rangle$ should be written as a linear combination of the superconformal Ishibashi states. Moreover the consistency of the rectangular torus with two parallel defects implies the Cardy condition. So we can work in $D(\textmd{CFT}_1 \otimes \textmd{CFT}_2)$ as far as the Cardy condition concerns.

This prescription may not be perfect. As we will see, the complete classification of the superconformal boundary in $D(\textmd{CFT}_1 \otimes \textmd{CFT}_2)$ is not a complete classification of the defects when both of CFT$_1$ and CFT$_2$ are the tricritical Ising model. We will require that the ``no defect'' boundary state is included and find a set of the boundary states which satisfies the Cardy condition. This set of the boundary states may not be a consistent set of the boundary states; for example, they may not satisfy the consistency condition of the bulk-boundary OPE in $D(\textmd{CFT}_1 \otimes \textmd{CFT}_2)$. The safest way to consider this OPE condition should be to deal with operators in unfolded picture.

If the folded theory $\textmd{CFT}_1\otimes \textmd{CFT}_2$ (or $D(\textmd{CFT}_1 \otimes \textmd{CFT}_2)$) is a rational conformal field theory, we can effectively use the boundary state formalism and expect to get some classifications of the boundary states (or defects in unfolded picture).
In the next subsection we just give a quick review of the boundary states in a rational conformal field theory.

\subsection{(Super)conformal boundary states and the Cardy condition}
Consider a rational conformal field theory with chiral algebra ${\cal A}$ (in our case, superconformal symmetry), and the boundary condition which preserve ${\cal A}$ out of ${\cal A}\otimes{\cal A}$. A simple way to preserve this chiral algebra is to impose the gluing conditions for the spin $s$ currents $J$ of ${\cal A}$ at the boundary on real axis as
\begin{align}
 J(z)= \bar{J}(\bar{z})|_{z=\bar{z}}. \label{general-glue}
 \end{align}
The upper-half plane is mapped into infinitely long strip with width $L$ by the transformation $w=t+i\sigma = {L \over \pi}\log z$. Imposing the periodic
boundary conditions in the $t$ direction, $t \sim t+T$, the strip becomes finite cylinder of circumference $T$ and length $L$. Some boundary conditions
$a$ and $b$ are imposed on the two boundaries, $\sigma=0$ and $\sigma=L$, respectively.

The partition function $Z_{a|b}$ in the cylinder can be calculated in two ways. Firstly consider $t$ as time-direction.
Then the cylinder can be interpreted as worldsheet of open string propagating $t$-direction.
The Hamiltonian is given in terms of a Virasoro generator in $z$-plane, $\frac{\pi}{L}(L_0 - {c \over 24})$. The open string Hilbert space, denoted by $\Hcal_{a,b}$,
 can be decomposed into the irreducible representations of the single chiral algebra ${\cal A}$ since the boundary condition preserves ${\cal A}$.  In other words there exist non-negative integers $n^{i}_{a,b}$ and $\Hcal_{a,b}$ is written as
\begin{align}
\Hcal_{a,b}= \bigoplus_{i} n^{i}_{a,b} \Hcal_i,
\end{align}
where $\Hcal_i$ are irreducible ${\cal A}$-modules.
The partition function can be written using the moduli parameter $q:=e^{2\pi i \tau}$, $\tau:= {i T \over 2 L}$ as
\begin{align}
Z_{a|b} =  \Tr_{\Hcal_{a,b}}e^{-\pi {T \over L} (L_0 - {c \over 24})} =  \Tr_{\Hcal_{a,b}} q^{L_0 - {c \over 24}} = \sum_{i} n^{j}_{a,b} \chi_{j}(q),
\label{Zopen}
\end{align}
where $\chi_{j}(q)$ is the character of the representation $\Hcal_j$ defined as
\begin{align}
 \chi_{j}(q):=\Tr_{\Hcal_i} q^{L_0-\frac{c}{24}}.
\end{align}

Secondly if we consider $\sigma$ direction as time, the cylinder can be considered as worldsheet of closed string propagating $\sigma$-direction. The cylinder is mapped into an annulus in $\xi$-plane by the transformation $\xi= \exp (-2\pi i w / T)$.
In this interpretation, boundaries $a,b$ become initial and final states, $|a\rangle$ and $|b\rangle$ respectively. These boundary states $|a\rangle, |b\rangle$ should satisfy
the gluing conditions \eqref{general-glue} which becomes
\begin{align}
\left( J_{n}- (-1)^{s} \bar{J}_{-n} \right)|a\rangle = 0,  \label{glue-closed}
\end{align}
in $\xi$-plane.
Solutions to \eqref{glue-closed} are spanned by special states called Ishibashi states\cite{Ishibashi:1988kg,Onogi:1988qk}. Let $\Hcal$ be the closed string Hilbert space. Then the Ishibashi states in $\Hcal$ are
\begin{align}
\{|j\rangle\rangle = \sum_{N} |(j,N) \rangle  \otimes \overline{U |(j,N)}\rangle    :   \Hcal_j \otimes \overline{\Hcal}_j \subset \Hcal\},
\end{align}
where $|j;0\rangle$ is a highest weight state and $|j;N\rangle$ are orthonormal basis of $\Hcal_j$. And the anti-unitary operator $U$ is defined by
\begin{align}
U\overline{|j;0\rangle} = \overline{|j;0\rangle }^* , \quad U\bar{J}_n = (-1)^{s} \bar{J}_n U , \quad \text{for any  $\bar{J}_{n}$.}
\end{align}
The closed string Hamiltonian is given by $\frac{2\pi}{T}(L_0 + \bar{L}_0 - {c \over 12})$ in $\xi$-plane. The partition function can be written as
\begin{align}
Z_{a|b} = \langle a | \tilde{q}^{{1 \over 2}(L_0 + \bar{L}_0 - {c \over 12})} |b\rangle,
\end{align}
where $\tilde{q} =e^{2\pi i \tilde{\tau}}$  and $\tilde{\tau}= - {1 \over \tau}$.
If we express two boundary states $|a\rangle,|b\rangle$ as linear combinations of the Ishibashi states,
\begin{align}
|a\rangle = \sum_{j} c_{a}^{j} |j\rangle\rangle, \quad |b\rangle = \sum_{j} c_{b}^{j} |j\rangle\rangle,
\end{align}
then the partition function can be expressed as
\begin{align}
Z_{a|b} &= \sum_{j} c_{a}^{j*} c_{b}^{j}\chi_{j} (\tilde{q}) = \sum_{j,j'} c_{a}^{j*} c_{b}^{j} S^{j'}_{j}\chi_{j'} (q).\label{Zclosed}
\end{align}
Here we used the fact
\begin{align}
\langle \langle j | \tilde{q}^{{1 \over 2}(L_0 + \bar{L}_0 - {c \over 12})} |j' \rangle \rangle = \delta_{j,j'}\chi_j (\tilde{q}),
\end{align}
and the modular transformation rule for the character $\chi_j $
\begin{align}
\chi_j (\tilde{q}) = \sum_{j'} S_{j}^{j'}\chi_{j'} (q ).
\end{align}

For the consistency of the theory, $Z_{a|b}$ calculated in two different ways must be identical.
If we assume the characters are linearly independent, eq.\eqref{Zopen} and eq.\eqref{Zclosed} imply
\begin{align}
n^{j'}_{a,b}=\sum_{j} c_{a}^{j*} c_{b}^{j} S^{j'}_{j}.\label{Cardy condition}
\end{align}
This equation gives non-trivial condition on the coefficients $c_{a}^{j}$ since $n^{j'}_{a,b}$ are non-negative integers; this is called Cardy condition.

Actually the characters are not linearly independent in the auxiliary theory in this paper. There are linear relations among $m=10$ superconformal characters
\begin{align}
\chi^{(10)}_{1,1,5} + \chi^{(10)}_{1,1,11} &= \chi^{(10)}_{1,3,1} + \chi^{(10)}_{1,3,7},  \nonumber
\\
\chi^{(10)}_{r,2,1} + \chi^{(10)}_{r,2,7} &= \chi^{(10)}_{r,3,4} + \chi^{(10)}_{r,3,8}, \quad r=2,4,6,8.
\end{align}
Thus eq.\eqref{Cardy condition} is not necessary for the equality of eq.\eqref{Zopen} and eq.\eqref{Zclosed} , and the Cardy condition is a little bit relaxed. But still the equality of eq.\eqref{Zopen} and eq.\eqref{Zclosed} gives a non-trivial constraint.

\section{Superconformal defects in the tricritical model}
\label{sec-main}

\subsection{Tricritical Ising model and its folded theory}
The tricritical Ising model is the first model ($m=3$ in eq.\eqref{1c}) of the $\Ncal=1$ unitary minimal series.  Its central charge is $c= \frac{7}{10}$. The toroidal partition function of this model is written as
\begin{align}
Z_{\textrm{tri}} = \frac{1}{2} \sum_{r,t,s} |\chi^{(m=3)}_{r,t,s}|^2,
\label{Ztri}
\end{align}
where $\chi^{(m=3)}_{r,t,s}$ are characters whose explicit forms are written in appendix \ref{app-minimal}.

Consider the product theory of two tricritical Ising models with spin structure aligned, denoted by $D(\textrm{tri}\otimes \textrm{tri})$. Its toroidal partition function is expressed as
\begin{align}
 Z_{D(\textrm{tri} \otimes \textrm{tri})} =& \frac{1}{4} \sum_{r_1 , s_1 ,r_2 ,s_2 \in \textrm{NS}} \big{[} |\chi^{(3)}_{r_1 , 1 ,s_1} \chi^{(3)}_{r_2 ,1, s_2} +\chi^{(3)}_{r_1 , 3, s_1} \chi^{(3)}_{r_2, 3, s_3} |^2
 +|\chi^{(3)}_{r_1,1,s_1} \chi^{(3)}_{r_2 ,3, s_2} + \chi^{(3)}_{r_1 ,3,s_1} \chi^{(3)}_{r_2, 1, s_2} |^2 \big{]}
\nonumber
\\
&
+ \frac{1}{4} \sum_{r_1 , s_1 , r_2 , s_2 \in \textrm{R}} 2 |\chi^{(3)}_{r_1, 2, s_1 } \chi^{(3)}_{r_2 , 2, s_2}|^2.\label{tritri}
\end{align}
Note that $D(\textrm{tri}\otimes \textrm{tri})$ is different from the naive direct product of two tricritical Ising model, which is denoted by $\textrm{tri}\otimes \textrm{tri}$.

Actually $D(\textrm{tri}\otimes \textrm{tri})$ is also an $\Ncal=1$ minimal model.  The central charge is $c=\frac{7}{5}$ ($m=10$).  Therefore the tensor product of two representations of NS (or R) algebra with $c=7/10$ $(m=3)$ can be decomposed into the representations of $c=7/5$ $(m=10)$ algebra. This decomposition can be explicitly seen from the character relations as follows.
\begin{align}
\chi^{(3)}_{1,1,1} \chi^{(3)}_{1,1,1} + \chi^{(3)}_{1,3,1}\chi^{(3)}_{1,3,1} &= \chi^{(10)}_{1,1,1} +\chi^{(10)}_{9,3,5} + \chi^{(10)}_{9,1,1} +\chi^{(10)}_{1,3,5},   \nonumber
\\
\chi^{(3)}_{1,1,3}\chi^{(3)}_{1,1,3} + \chi^{(3)}_{1,3,3}\chi^{(3)}_{1,3,3} &= \chi^{(10)}_{3,1,1}+\chi^{(10)}_{7,3,5}+\chi^{(10)}_{7,1,1}+\chi^{(10)}_{3,3,5}, \nonumber
\\
\chi^{(3)}_{1,1,3}\chi^{(3)}_{1,3,3} + \chi^{(3)}_{1,3,3}\chi^{(3)}_{1,1,3} &= \chi^{(10)}_{3,1,5}+\chi^{(10)}_{7,3,1}+\chi^{(10)}_{7,1,5}+\chi^{(10)}_{3,3,1}, \nonumber
\\
\chi^{(3)}_{1,1,1} \chi^{(3)}_{1,3,1} + \chi^{(3)}_{1,3,1}\chi^{(3)}_{1,1,1}& = \chi^{(10)}_{1,1,5} +\chi^{(10)}_{9,3,1} + \chi^{(10)}_{9,1,5} + \chi^{(10)}_{1,3,1},  \nonumber
\\
\chi^{(3)}_{1,1,1}\chi^{(3)}_{1,1,3} + \chi^{(3)}_{1,3,1}\chi^{(3)}_{1,3,3} &= \chi^{(10)}_{5,1,1}+\chi^{(10)}_{5,3,5}, \nonumber
\\
\chi^{(3)}_{1,1,1}\chi^{(3)}_{1,3,3} + \chi^{(3)}_{1,3,1}\chi^{(3)}_{1,1,3} &= \chi^{(10)}_{5,1,5}+\chi^{(10)}_{5,3,1}, \nonumber
\\
\chi^{(3)}_{1,2,4}\chi^{(3)}_{1,2,4} &= \chi^{(10)}_{1,2,4} +\chi^{(10)}_{9,2,4}, \nonumber
\\
\chi^{(3)}_{1,2,2}\chi^{(3)}_{1,2,2} & = \chi^{(10)}_{3,2,4} +\chi^{(10)}_{7,2,4}, \nonumber
\\
\chi^{(3)}_{1,2,4}\chi^{(3)}_{1,2,2} & = \chi^{(10)}_{5,2,4}.
\label{char-identity}
\end{align}
These relations are checked using the explicit form of the characters \eqref{character-formula0}-\eqref{character-formula1} by $q$ expansion.

One can rewrite the partition function \eqref{tritri} in terms of $m=10$ characters using \eqref{char-identity} as
\begin{align}
 Z_{D(\textrm{tri} \otimes \textrm{tri})}=&
\sum_{t=1,3}\big[|\chi^{(10)}_{1,t,1}+\chi^{(10)}_{1,t,7}+\chi^{(10)}_{9,t,1}+\chi^{(10)}_{9,t,7}|^2+|\chi^{(10)}_{3,t,1}+\chi^{(10)}_{3,t,7}+\chi^{(10)}_{7,t,1}+\chi^{(10)}_{7,t,7}|^2
\nonumber\\
&+2|\chi^{(10)}_{5,t,1}+\chi^{(10)}_{5,t,7}|^2\big]
+2|\chi^{(10)}_{1,2,4}+\chi^{(10)}_{1,2,8}|^2
+2|\chi^{(10)}_{3,2,4}+\chi^{(10)}_{3,2,8}|^2
+4|\chi^{(10)}_{5,2,4}|^2
\nonumber\\
=& \frac{1}{2} \sum_{\substack{r+s+t=\text{odd}\\
 \bar{r}+\bar{s}+t=\text{odd}}} N^{D_6}_{r\bar{r}} N^{E_6}_{s\bar{s}} \chi^{(10)}_{r,t,s} \bar{\chi}^{(10)}_{\bar{r},t , \bar{s}},
\end{align}
where $N^{D_6}_{r,\bar{r}}$ and  $N^{E_6}_{s,\bar{s}}$ stand for $D_6$ -type modular invariant of $\widehat{SU}(2)_{8}$ and $E_6$ -type modular invariant of $\widehat{SU}(2)_{10}$ respectively. Therefore we conclude that the tensor product with aligned spin structure $D(\textrm{tri} \otimes \textrm{tri})$ is $c=7/5$, $D_6-E_6$ theory.

We can classify the boundary states in this theory following \cite{Cardy:1989ir,Behrend:1999bn}.  The result is written in appendix \ref{app-m=10}.
We obtained two distinct complete sets of boundary states.
However neither of these two sets can be interpreted as a complete set of the superconformal defects in the tricritical Ising model.  This is because they do not include ``no defect.''  Another problem is that $D(\textrm{tri} \otimes \textrm{tri})$ includes twice as many Ramond states as unfolded theory as seen in eq.\eqref{tritri}. When unfolding, a state in $D(\textrm{tri} \otimes \textrm{tri})$ are mapped to an operator in the tricritical Ising model.  If the boundary states include both $t=2$ and $t=\tilde 2$ Ramond states, it is impossible to map those states to operators in the tricritical Ising model while keeping the super Virasoro action and inner product structure.

In the next subsection, we propose a set of boundary states which are free from these problems.
\subsection{18 superconformal defects in the tricritical Ising model.}
In this subsection, we consider the set of the boundary states which satisfies the following criteria.
\begin{enumerate}
 \item It includes ``no defect.''
 \item It satisfies the Cardy condition.
\end{enumerate}
We find a set of 18 boundary states which satisfies the above criteria. This is the main result of this paper.

Let us first explain ``no defect'' boundary state.
Boundary state $|N\rangle$ in $D(\textrm{tri} \otimes \textrm{tri})$ which corresponds to ``no defect'' in the tricritical Ising model can be chosen as (see Appendix \ref{app-m=10} for notation used in this section)
\begin{align}
|N\rangle  =& \frac{1}{2} |2,6;\widetilde{NS} \rangle + \frac{1}{\sqrt{2}} |2,1;NS \rangle + |2,1;R \rangle  \nonumber
\\
=&|(1,1,1)_{10} \rangle \rangle + |(1,3,5)_{10} \rangle\rangle +|(1,2,4)_{10}\rangle \rangle +
|(3,1,1)_{10} \rangle \rangle + |(3,3,5)_{10} \rangle\rangle +|(3,2,4)_{10} \rangle \rangle  \nonumber
\\
&- |(7,1,1)_{10} \rangle \rangle - |(7,3,5)_{10} \rangle\rangle -|(7,2,4)_{10} \rangle \rangle -
 |(9,1,1)_{10} \rangle \rangle - |(9,3,5)_{10} \rangle\rangle -|(9,2,4)_{10} \rangle \rangle.
\end{align}
Presence of ``no defect'' has the same effects as absence of defects. Therefore the annulus amplitude of two ``no defect'' boundary condition is equal to the toroidal partition function of the unfolded theory i.e. the tricritical Ising model. Actually it can be checked explicitly that
\begin{align}
\langle N | \tilde{q}^{{1 \over 2}(L_0 + \bar{L}_0 - {c \over 12})} |N\rangle = Z_{\textrm{tri}}(\tilde{q})=Z_{\textrm{tri}}(q),
\end{align}
is satisfied with $Z_{\textrm{tri}}$ of eq.\eqref{Ztri}. In section \ref{sec:transmission}, we check that the transmission coefficient $\mathcal{T}$ for $|N\rangle$ equals to 1. Moreover $|N\rangle$ can be expanded as (see \eqref{nodefect_expansion} and \eqref{permutation_Ishibashi})
\begin{align}
|N\rangle = \sum_{|a\rangle \in \mathcal{H}_{\textrm{tri}}} |a\rangle \otimes |a\rangle .
\end{align}
This states corresponds to identity operator $I$ before the folding.
\begin{align}
I = \sum_{|a\rangle \in \mathcal{H}_{tri}} |a\rangle \langle a|.
\end{align}
Note that ``no defect'' is not consistent (in the sense of Cardy condition) with 36 boundary states in $D(\textrm{tri}\otimes \textrm{tri})$ (see Appendix \ref{app-m=10}).

Next let us consider the set of boundary states which includes the above ``no defect'' boundary state and satisfies the Cardy condition.  We found the following 18 boundary states which meet the criteria.
\begin{align}
\textrm{$A_{\pm}$-type : } | (a,b)\rangle _{A_{\pm }}&= {1 \over 2}|a,b;\widetilde{NS}\rangle  + {1 \over \sqrt{2}} |a,\rho(b);NS \rangle \pm |a,\rho (b);R \rangle, \nonumber
\\
& \quad \textrm{where } (a,b)=(1,3),(3,3),(5,3),(6,3),(2,6),(4,6),
  \nonumber
\\
\textrm{$B$-type : } |(a,b)\rangle_B  &= |a,b;\widetilde{NS}\rangle  + {1 \over \sqrt{2}} |a,\rho^{-1}(b); NS \rangle, \nonumber
\\
& \quad \textrm{where }(a,b)=(1,1),(3,1),(5,1),(6,1),(2,2),(4,2).\label{18defects}
\end{align}
Here we use the function $\rho$ defined in eq.\eqref{def-rho}.
Actually one can identify $|(2,6)\rangle_{A+}$ as the ``no defect'' boundary state. On the other hand, the Cardy condition can be checked as follows. Let us define the number $n^{r,t,s}_{(a,b,X),(a',b',Y)},\ (X,Y=A_{\pm},B)$ as
\begin{align}
 {}_X\langle (a,b)|\tilde{q}^{\frac12(L_0+\bar{L}_{0}-c/12)}|(a',b')\rangle_{Y}
=\sum_{[r,t,s]}n^{r,t,s}_{(a,b,X),(a',b',Y)}\chi^{(10)}_{r,t,s}(q).
\end{align}
These coefficients are calculated
 by using \eqref{overlap1} and the relations \eqref{n-relation0}--\eqref{n-relation1} as
\begin{align}
&n^{r,1,s}_{(a,b,A_{\pm}) , (a',b',A_{\pm})} = n^{r}_{a,a'}(D_6) n^{s}_{\rho(b),\rho(b')}(E_6),
\nonumber
\\
&n^{r,2,s}_{(a,b,A_{\pm}) , (a',b',A_{\pm})} = n^{r}_{a,a'}(D_6) n^{s}_{b,\rho(b')}(E_6), \nonumber
\\
&n^{r,t,s}_{(a,b,A_{\pm}) , (a',b',A_{\mp})} =n^{r,4-t,s}_{(a,b,A_{\pm}) , (a',b',A_{\pm})}, \nonumber
\\
&n^{r,1,s}_{(a,b,A_{\pm}), (a',b',B)} = n^{r}_{a,a'}(D_6) n^{s}_{b,b'}(E_6), \nonumber
\\
&n^{r,2,s}_{(a,b,A_{\pm}),(a',b',B)} = n^{r}_{a,a'}(D_6) n^{s}_{b,\rho^{-1}(b')}(E_6), \nonumber
\\
&n^{r,1,s}_{(a,b,B),(a',b',B)} = n^{r}_{a,a'}(D_6) n^{s}_{\rho^{-1}(b), \rho^{-1}(b')}(E_6),
\nonumber
\\
&n^{r,2,s}_{(a,b,B),(a',b',B)} = 2 n^{r}_{a,a'}(D_6) n^{s}_{b,\rho^{-1}(b')}(E_6),
\nonumber
\\
&n^{r,3,s}_{(a,b,X),(a',b',Y)}=n^{10-r,1,12-s}_{(a,b,X),(a',b',Y)}.
\end{align}
As a result all the coefficients $n^{r,t,s}_{(a,b,X),(a',b',Y)}$ are shown to be non-negative integers, namely, the Cardy condition is satisfied.
Therefore we conclude that this set of the boundary states meet the criteria.

It seems that this set of 18 boundary states is the maximal one which meets the criteria.

In this paper, we did not consider the consistency of the bulk-boundary OPE. The 18 boundary states we obtained here may not be consistent in OPE in the bulk theory $D(\textrm{tri} \otimes \textrm{tri})$, since there are other two sets of 36 boundary states which satisfy the Cardy condition. However this is not a problem because what we want to do is to obtain the consistent set of defects in tricritical Ising model, and it is different from the consistent boundary of $D(\textrm{tri} \otimes \textrm{tri})$. Hence, OPE consistency should be checked in the unfolded theory (tricritical Ising model) instead of $D(\textrm{tri} \otimes \textrm{tri})$. It is an interesting future problem.

\subsection{Reflection and transmission coefficients}
\label{sec:transmission}
Let us now consider reflection and transmission coefficients $\mathcal{R},\mathcal{T}$ for superconformal defects obtained above. Reflection and transmission coefficients are defined and investigated in \cite{Quella:2006de}.  These coefficients in general are defined as follows.  First, we consider the matrix $R_{ij}$ for a boundary state $|B\rangle$ defined as
\begin{align}
 R_{ij}=\frac{\langle 0|L_{2}^{(i)}\bar{L}^{(j)}_{2}|B\rangle}{\langle 0|B\rangle}.
\end{align}
Then, the reflection and transmission coefficients $\mathcal{R},\mathcal{T}$ are defined as
\begin{align}
 \mathcal{R}:=\frac{2}{c_1+c_2}(R_{11}+R_{22}),\qquad
 \mathcal{T}:=\frac{2}{c_1+c_2}(R_{12}+R_{21}),
\end{align}
where $c_1$ and $c_2$ are the central charges of $CFT_1$ and $CFT_2$ respectively. In our problem, $c_1=c_2=7/10$. These coefficients satisfy the relation $\mathcal{R}+\mathcal{T}=1$, so we will only write down $\mathcal{T}$.

 These coefficients for each defect in \eqref{18defects} can be calculated as
\begin{align}
\mathcal{T}=1 \quad &: \quad |(2,6)\rangle_{A_{\pm}}, \quad |(4,6)\rangle_{A_{\pm}}, \nonumber
\\
\mathcal{T}=0 \quad &: \quad |(1,1)\rangle_B, \quad |(3,1)\rangle_B , \quad |(5,1)\rangle_B , \quad |(6,1)\rangle_B, \nonumber
\\
\mathcal{T}=\frac{3}{3+\sqrt{3}} \quad &: \quad |(1,3)\rangle_{A_\pm} , \quad |(3,3)\rangle_{A_\pm}, \quad |(5,3)\rangle_{A_\pm}, \quad |(6,3)\rangle_{A_\pm},  \nonumber
\\
\mathcal{T}=\frac{\sqrt{3}}{3+\sqrt{3}}\quad &: \quad |(2,2)\rangle_B , \quad |(4,2)\rangle_B.\label{transmission}
\end{align}
So this set includes totally transmitting and totally reflecting defects as well as intermediate ones.

Totally reflecting ($\mathcal{T}=0$) and totally transmitting ($\mathcal{T}=1$) defects can be expressed in terms of factorized Ishibashi states and permutation ones, respectively, of the tensor product of the two tricritical Ising models. Factorized Ishibashi states can be expressed as
\begin{align}
&|(r,t,s,r',t',s')_3 \rangle\rangle_R \nonumber
\\
&= \sum_{N,M} \big{[} |(r,t,s)_3,N \rangle \otimes U |(r,t,s)_3,N\rangle \big{]}^{(1)} \otimes \big{[} |(r',t',s')_3  ,M \rangle \otimes U |(r',t',s')_3,M\rangle \big{]}^{(2)},
\end{align}
while permutation Ishibashi states are written as
\begin{align}
&|(r,t,s,r,t,s)_3 \rangle\rangle_T \nonumber
\\
&= \sum_{N,M} \big{[} |(r,t,s)_3,N \rangle \otimes U |(r,t,s)_3,M\rangle \big{]}^{(1)} \otimes \big{[} |(r,t,s)_3  ,M \rangle \otimes U |(r,t,s)_3,N\rangle \big{]}^{(2)}. \label{permutation_Ishibashi}
\end{align}
The overlaps among these Ishibashi states become
\begin{align}
& _R \langle \langle (r_1,t_1,s_1,r'_1,t'_1,s'_1)_3 | \tilde{q}^{{1 \over 2}(L_0 + \bar{L}_0 - {c \over 12})} | (r_2 , t_2 , s_2 , r'_2, t'_2 , s'_2)_3 \rangle\rangle_R  \nonumber
\\
&\qquad\qquad\qquad\qquad = \delta_{[r_1 ,t_1 ,s_1],[r_2,t_2,s_2]} \delta_{[r'_1 ,t'_1 ,s'_1],[r'_2,t'_2,s'_2]} \chi^{(3)}_{r_1,t_1,s_1}(\tilde{q}) \chi^{(3)}_{r'_1 ,t'_1, s'_1 } (\tilde{q}),
\\ \nonumber
&_T \langle \langle (r_1,t_1,s_1,r_1,t_1,s_1)_3 |
 \tilde{q}^{{1 \over 2}(L_0 + \bar{L}_0 - {c \over 12})}
 | (r_2 , t_2 , s_2 , r_2, t_2 , s_2)_3 \rangle\rangle_T
 \nonumber\\&\qquad\qquad\qquad\qquad
= \delta_{[r_1 ,t_1 ,s_1],[r_2,t_2,s_2]}
 \chi^{(3)}_{r_1,t_1,s_1}(\tilde{q}) \chi^{(3)}_{r'_1 ,t'_1, s'_1 } (\tilde{q}),
\\
& _T \langle \langle (r,t,s,r,t,s)_3 |\tilde{q}^{{1 \over 2}(L_0 + \bar{L}_0 - {c \over 12})} | (r , t, s , r, t , s)_3 \rangle\rangle_R =\chi^{(3)}_{r,t,s}(\tilde{q}^2).
\end{align}
Some linear combination of these Ishibashi states can be expressed in terms of Ishibashi states in the $D_6 -E_6$ theory.
\begin{align}
& |(1,1,1,1,1,1)_3 \rangle \rangle_T +|(1,3,1,1,3,1)_3 \rangle\rangle_T \nonumber\\&\qquad\qquad\qquad\qquad
=|(1,1,1)_{10}\rangle\rangle  - |(9,3,5)_{10}\rangle \rangle - |(9,1,1)_{10}\rangle \rangle + |(1,3,5)_{10}\rangle\rangle, \nonumber
\\
& |(1,1,3,1,1,3)_3 \rangle \rangle_T +|(1,3,3,1,3,3)_3 \rangle\rangle_T
\nonumber\\&\qquad\qquad\qquad\qquad
=|(3,1,1)_{10}\rangle\rangle  - |(7,3,5)_{10}\rangle \rangle - |(7,1,1)_{10}\rangle \rangle + |(3,3,5)_{10}\rangle\rangle,
\nonumber\\&
|(1,1,1,1,1,1)_3 \rangle \rangle_R - |(1,3,1,1,3,1)_3 \rangle\rangle_R
 \nonumber\\&\qquad\qquad\qquad\qquad
=|(1,1,1)_{10}\rangle\rangle  - |(9,3,5)_{10}\rangle \rangle + |(9,1,1)_{10}\rangle \rangle - |(1,3,5)_{10}\rangle\rangle, \nonumber
\\
& |(1,1,3,1,1,3)_3 \rangle \rangle_R -|(1,3,3,1,3,3)_3 \rangle\rangle_R \nonumber\\&\qquad\qquad\qquad\qquad
=|(3,1,1)_{10}\rangle\rangle  - |(7,3,5)_{10}\rangle \rangle + |(7,1,1)_{10}\rangle \rangle - |(3,3,5)_{10}\rangle\rangle, \nonumber
\\
&|(1,2,4,1,2,4)_3\rangle\rangle_T =|(1,2,4)_{10}\rangle \rangle - |(9,2,4)_{10}\rangle \rangle,   \nonumber
\\
&|(1,2,2,1,2,2)_3\rangle\rangle_T  =|(3,2,4)_{10}\rangle \rangle - |(7,2,4)_{10}\rangle \rangle, \nonumber
\\
&|(1,1,1,1,1,3)_3 \rangle \rangle_R - |(1,3,1,1,3,3)_3\rangle\rangle_R =|(5,1,1)_{10}\rangle\rangle - |(5,3,5)_{10}\rangle\rangle,   \nonumber
\\
& |(1,1,3,1,1,1)_3 \rangle \rangle_R - |(1,3,3,1,3,1)_3\rangle\rangle_R =|(5',1,1)_{10}\rangle\rangle - |(5',3,5)_{10}\rangle\rangle.
\label{Ish:m=3,m=10}
\end{align}
This identification can be justified by comparing the overlaps among these states in both expressions. This can be checked using the character identities \eqref{char-identity} and the followings.
\begin{align}
(\chi^{(10)}_{1,1,1} + \chi^{(10)}_{9,3,5} - \chi^{(10)}_{9,1,1} - \chi^{(10)}_{1,3,5} )(\tilde{q}) = ( \chi^{(3)}_{1,1,1} - \chi^{(3)}_{1,3,1} )({\tilde{q}^2}), \nonumber
\\
(\chi^{(10)}_{3,1,1} + \chi^{(10)}_{7,3,5} - \chi^{(10)}_{7,1,1} - \chi^{(10)}_{3,3,5} )(\tilde{q}) = ( \chi^{(3)}_{1,1,3} - \chi^{(3)}_{1,3,3} )({\tilde{q}^2}).
\end{align}
It was checked up to some order by the $q$ expansion.

Now we will give the explicit expression of totally transmitting (reflecting) defects in terms of permutation (factorizing) Ishibashi states using \eqref{Ish:m=3,m=10}\footnote{Note that if we expand a totally transmitting (reflecting) defect in terms of Ishibashi states in the $D_6-E_6$ theory, only the linear combinations those can be expressed via \eqref{Ish:m=3,m=10} in terms of permutation (factorizing) Ishibashi states appear.  }. For totally transmitting defects,
\begin{align}
|(2,6)\rangle_{A_\pm} &
= |(1,1,1,1,1,1)_3 \rangle \rangle_T+ |(1,3,1,1,3,1)_3 \rangle\rangle_T  \nonumber
\\
& +|(1,1,3,1,1,3)_3 \rangle\rangle_T+ |(1,3,3,1,3,3)_3 \rangle\rangle_T \nonumber
\\
&\pm \big{[}|(1,2,4,1,2,4)_3 \rangle\rangle_T + |(1,2,2,1,2,2)_3\rangle \rangle_T \big{]}, \label{nodefect_expansion}
\\
|(4,6)\rangle_{A_\pm} &= \sqrt{20}\alpha_+ \big{[} |(1,1,1,1,1,1)_3 \rangle \rangle_T + |(1,3,1,1,3,1)_3 \rangle\rangle_T \big{]}  \nonumber
\\
&- \sqrt{20} \alpha_- \big{[} |(1,1,3,1,1,3)_3 \rangle\rangle_T+ |(1,3,3,1,3,3)_3 \rangle\rangle_T \big{]} \nonumber
\\
&\pm \big{[} \sqrt{20} \alpha_+ |(1,2,4,1,2,4)_3 \rangle\rangle_T - \sqrt{20} \alpha_- |(1,2,2,1,2,2)_3 \rangle \rangle_T \big{]},
\end{align}
where $\alpha_{\pm}:=\frac{5\pm \sqrt{5}}{20}$. For totally reflecting defects,
\begin{align}
|(1,1)\rangle_B &= 2 \sqrt{\alpha_-} \big{[}|(1,1,1,1,1,1)_3 \rangle\rangle_R - |(1,3,1,1,3,1)_3 \rangle\rangle_R \big{]} \nonumber
\\
&+ 2\sqrt{\alpha_+} \big{[}|(1,1,3,1,1,3)_3 \rangle\rangle_R - |(1,3,3,1,3,3)_3 \rangle\rangle_R \big{]}  \nonumber
\\
&+  5^{-1/4}\sqrt{2} \big{[}|(1,1,1,1,1,3)_3 \rangle\rangle_R -|(1,3,1,1,3,3)_3 \rangle\rangle_R  \big{]}   \nonumber
\\
&+  5^{-1/4}\sqrt{2} \big{[}|(1,1,3,1,1,1)_3 \rangle\rangle_R -|(1,3,3,1,3,1)_3 \rangle\rangle_R  \big{]},
\end{align}
\begin{align}
|(3,1)\rangle_B &= 2 \sqrt{20 \alpha_+^3} \big{[}|(1,1,1,1,1,1)_3 \rangle\rangle_R - |(1,3,1,1,3,1)_3 \rangle\rangle_R \big{]} \nonumber
\\
&+ 2\sqrt{20 \alpha_-^3} \big{[}|(1,1,3,1,1,3)_3 \rangle\rangle_R - |(1,3,3,1,3,3)_3 \rangle\rangle_R \big{]}  \nonumber
\\
&-  5^{-1/4}\sqrt{2} \big{[}|(1,1,1,1,1,3)_3 \rangle\rangle_R -|(1,3,1,1,3,3)_3 \rangle\rangle_R  \big{]}   \nonumber
\\
&-  5^{-1/4}\sqrt{2} \big{[}|(1,1,3,1,1,1)_3 \rangle\rangle_R -|(1,3,3,1,3,1)_3 \rangle\rangle_R  \big{]},
\end{align}
\begin{align}
|(5,1)\rangle_B &= 2 \sqrt{\alpha_+} \big{[}|(1,1,1,1,1,1)_3 \rangle\rangle_R - |(1,3,1,1,3,1)_3 \rangle\rangle_R \big{]}\nonumber
\\
&- 2\sqrt{\alpha_-} \big{[}|(1,1,3,1,1,3)_3 \rangle\rangle_R - |(1,3,3,1,3,3)_3 \rangle\rangle_R \big{]}  \nonumber
\\
&+  2^{3/2}5^{1/4} \alpha_+  \big{[}|(1,1,1,1,1,3)_3 \rangle\rangle_R -|(1,3,1,1,3,3)_3 \rangle\rangle_R  \big{]}   \nonumber
\\
&-  2^{3/2}5^{1/4} \alpha_-  \big{[}|(1,1,3,1,1,1)_3 \rangle\rangle_R -|(1,3,3,1,3,1)_3 \rangle\rangle_R  \big{]},
\end{align}
\begin{align}
|(6,1)\rangle_B &= 2 \sqrt{\alpha_+} \big{[}|(1,1,1,1,1,1)_3 \rangle\rangle_R - |(1,3,1,1,3,1)_3 \rangle\rangle_R \big{]} \nonumber
\\
& -2\sqrt{\alpha_-} \big{[}|(1,1,3,1,1,3)_3 \rangle\rangle_R - |(1,3,3,1,3,3)_3 \rangle\rangle_R \big{]}  \nonumber
\\
& -2^{3/2} 5^{1/4} \alpha_- \big{[}|(1,1,1,1,1,3)_3 \rangle\rangle_R -|(1,3,1,1,3,3)_3 \rangle\rangle_R  \big{]}   \nonumber
\\
& +2^{3/2} 5^{1/4} \alpha_+ \big{[}|(1,1,3,1,1,1)_3 \rangle\rangle_R -|(1,3,3,1,3,1)_3 \rangle\rangle_R  \big{]}.
\end{align}
These four totally reflecting defects can be expressed in terms of boundary states in tricritical Ising model. According to Cardy's prescription \cite{Cardy:1989ir}, there are 6 boundary states in tricritical Ising model labeled by weights.
\begin{align}
|0 \rangle &= 2^{-1/4}\alpha^{1/4}_- \big{[} |(1,1,1)\rangle \rangle +|(1,3,1)\rangle \rangle \big{]} + 2^{-1/4}\alpha^{1/4}_+\big{[}|(1,1,3)\rangle\rangle+|(1,3,3)\rangle \rangle \big{]} \nonumber
\\
&+\alpha^{1/4}_+ |(1,2,2)\rangle \rangle + \alpha^{1/4}_- |(1,2,4)\rangle \rangle , \nonumber
\\
|\frac{3}{2}\rangle &= 2^{-1/4}\alpha^{1/4}_- \big{[} |(1,1,1)\rangle \rangle +|(1,3,1)\rangle \rangle \big{]} + 2^{-1/4}\alpha^{1/4}_+\big{[}|(1,1,3)\rangle\rangle+|(1,3,3)\rangle \rangle \big{]} \nonumber
\\
&-\alpha^{1/4}_+ |(1,2,2)\rangle \rangle - \alpha^{1/4}_- |(1,2,4)\rangle \rangle , \nonumber
\\
|\frac{1}{10}\rangle &= 2^{-1/4}\alpha^{1/2}_+ \alpha^{-1/4}_- \big{[} |(1,1,1)\rangle \rangle +|(1,3,1)\rangle \rangle \big{]} - 2^{-1/4}\alpha^{1/2}_- \alpha^{-1/4}_+ \big{[}|(1,1,3)\rangle\rangle+|(1,3,3)\rangle \rangle \big{]} \nonumber
\\
&-\alpha^{1/2}_- \alpha^{-1/4}_+ |(1,2,2)\rangle \rangle + \alpha^{1/2}_+ \alpha^{-1/4}_- |(1,2,4)\rangle \rangle , \nonumber
\\
|\frac{3}{5}\rangle &= 2^{-1/4}\alpha^{1/2}_+ \alpha^{-1/4}_- \big{[} |(1,1,1)\rangle \rangle +|(1,3,1)\rangle \rangle \big{]} - 2^{-1/4}\alpha^{1/2}_- \alpha^{-1/4}_+ \big{[}|(1,1,3)\rangle\rangle+|(1,3,3)\rangle \rangle \big{]} \nonumber
\\
&+\alpha^{1/2}_- \alpha^{-1/4}_+ |(1,2,2)\rangle \rangle - \alpha^{1/2}_+ \alpha^{-1/4}_- |(1,2,4)\rangle \rangle , \nonumber
\\
|\frac{7}{16}\rangle & = 2^{1/4}\alpha^{1/4}_-  \big{[} |(1,1,1)\rangle \rangle -|(1,3,1)\rangle \rangle \big{]} + 2^{1/4}\alpha^{1/4}_+ \big{[}|(1,1,3)\rangle\rangle-|(1,3,3)\rangle \rangle \big{]} \nonumber
\\
|\frac{3}{80}\rangle & = 2^{1/4}\alpha^{1/2}_+ \alpha^{-1/4}_-  \big{[} |(1,1,1)\rangle \rangle -|(1,3,1)\rangle \rangle \big{]} - 2^{1/4}\alpha^{1/2}_- \alpha^{-1/4}_+ \big{[}|(1,1,3)\rangle\rangle-|(1,3,3)\rangle \rangle \big{]} 
\end{align} 
Here $|(r,t,s)\rangle\rangle$ are Ishibashi states in tricritical Ising model which is related to the factorized Ishibashi states as
\begin{align}
|(r,t,s,r',t',s')\rangle\rangle_R = |(r,t,s)\rangle \rangle \otimes |(r',t',s')\rangle\rangle. 
\end{align}
Four totally reflecting defects can be written as
\begin{align}
2|(1,1)\rangle_B &=(|0 \rangle + |\frac{3}{2}\rangle ) \otimes |\frac{7}{16}\rangle + |\frac{7}{16}\rangle \otimes (|0 \rangle + |\frac{3}{2}\rangle ), \nonumber
\\
2|(3,1)\rangle_B &=(|\frac{1}{10} \rangle + |\frac{3}{5}\rangle ) \otimes |\frac{3}{80}\rangle + |\frac{3}{80}\rangle \otimes (|\frac{1}{10} \rangle + |\frac{3}{5}\rangle ), \nonumber
\\
2|(5,1)\rangle_B &= (|\frac{1}{10} \rangle + |\frac{3}{5}\rangle ) \otimes |\frac{7}{16}\rangle + |\frac{3}{80}\rangle \otimes (|0 \rangle + |\frac{3}{2}\rangle ), \nonumber
\\
2|(6,1)\rangle_B &= (| 0  \rangle + |\frac{3}{2}\rangle ) \otimes |\frac{3}{80}\rangle + |\frac{7}{16}\rangle \otimes (|\frac{1}{10} \rangle + |\frac{3}{5}\rangle ) . 
\end{align}
Thus these totally reflecting defects $|(a,b)\rangle_B,\ (a,b)=(1,1),(3,1),(5,1),(6,1)$ cannot be expressed as linear combinations of the factorizing boundary states with non-negative integer coefficients, though twice of those states $2|(a,b)\rangle_B$ can.
One possible interpretation is that the consistent defects are $2|(a,b)\rangle_B$ instead $|(a,b)\rangle_B$ for $(a,b)=(1,1),(3,1),(5,1),(6,1)$.
Even if $|(a,b)\rangle_B$ are replaced by $2|(a,b)\rangle_B$, those 18 boundary states satisfy the Cardy condition.

\subsection*{Acknowledgments}
We would like to thank Changrim Ahn, Tohru Eguchi, Yasuaki Hikida, Kazuo Hosomichi, Shinsuke Kawai, Yu Nakayama, Soo-Jong Rey, Yuji Sugawara, and Tadashi Takayanagi for useful discussion and comments.  Discussions during ``Focus Program on Liouville, Integrability and Branes (4)'' at APCTP (Dec. 11-24, 2007), the YITP workshop YITP-W-08-04 on ``Development of Quantum Field Theory and String Theory,'' (Jul. 27- Aug. 1, 2008) and ``Summer Institute 2008'' at Yamanashi (Aug. 3-13, 2008) were useful to complete this work.
S.Y. was supported in part by KOFST BP Korea Program, KRF-2005-084-C00003, EU FP6 Marie Curie Research and Training Networks MRTN-CT-2004-512194 and HPRN-CT-2006-035863 through MOST/KICOS.

\appendix
\section{$\Ncal=1$ superconformal minimal models}
\label{app-minimal}
$\Ncal=1$ superconformal unitary minimal models are expressed by the coset model $\frac{\widehat{SU}(2)_{m-2} \otimes \widehat{SU}(2)_2}{\widehat{SU}(2)_{m}}$, $(m=3,4,5,\dots)$. The central charge is
\begin{align}
 c=\frac32\left(1-\frac{8}{m(m+2)}\right).\label{1c}
\end{align}

A module of this model is labeled by three integers $(r,t,s)$
\begin{align}
&  r=1,2,\dots,m-1,\qquad
 t=1,2,3,\qquad
 s=1,2,\dots,m+1,\\
& r+t+s=(\text{odd integer}),
\end{align}
under the identification
\begin{align}
 (r,t,s)\sim (m-r,4-t,m+2-s).\label{equiv}
\end{align}
The equivalence class of the equivalence relation \eqref{equiv} is denoted by $[r,t,s]$.
The module with $t=1$ or $3$ is in NS sector, while one with $t=2$ is in R sector.

The characters are denoted by
$\chi^{(m)}_{r,t,s}(q)$. These characters are explicitly written as follows\cite{Goddard:1986ee}. For NS sector ($r+s=$even)
\begin{align}
& \ch^{(m)}_{r,s}(q):=\chi^{(m)}_{r,1,s}(q)+\chi^{(m)}_{r,3,s}(q)
 =K^{(m)}_{r,s}(q)\; q^{-\frac{1}{16}}\prod_{n=1}^{\infty}\frac{1+q^{n-\frac12}}{1-q^{n}},\qquad q:=e^{2\pi i\tau},\label{character-formula0}\\
& \chtilde^{(m)}_{r,s}(q):=\chi^{(m)}_{r,1,s}(q)-\chi^{(m)}_{r,3,s}(q)
 =\widetilde{K}^{(m)}_{r,s}(q)\; q^{-\frac{1}{16}}\prod_{n=1}^{\infty}\frac{1-q^{n-\frac12}}{1-q^{n}},
\end{align}
while for R sector ($r+s=$odd)
\begin{align}
& \ch^{(m)}_{r,s}(q):=\chi^{(m)}_{r,2,s}(q)
 =K^{(m)}_{r,s}(q)\; \prod_{n=1}^{\infty}\frac{1+q^{n}}{1-q^{n}}.
\end{align}
Here the functions $K^{(m)}_{r,s}(q)$ and $\widetilde{K}^{(m)}_{r,s}(q)$ are defined as
\begin{align}
& K^{(m)}_{r,s}(q):=\sum_{n\in\Zb}\left(
q^{\Delta^{(m)}_{n,r,s}}-q^{\Delta^{(m)}_{n,r,-s}}\right),\qquad
\Delta^{(m)}_{n,r,s}:=\frac{[2m(m+2)n+ms-(m+2)r]^2}{8m(m+2)},\\
& \widetilde{K}^{(m)}_{r,s}(q):=\sum_{n\in\Zb}\left(
(-1)^{\frac{r-s}{2}+mn}q^{\Delta^{(m)}_{n,r,s}}-(-1)^{\frac{r+s}{2}+mn}q^{\Delta^{(m)}_{n,r,-s}}\right).
\label{character-formula1}
\end{align}

\section{Boundary states in $D_6-E_6$ theory}
\label{app-m=10}
\subsection{Boundary states in su(2) WZW model}
Here, we summarize some notations and facts on the boundary states in su(2) WZW model with ADE modular invariants, especially $D_6$ and $E_6$. For more detailed arguments, see \cite{Behrend:1999bn}.

Modular invariants of level $k$ su(2) WZW model are all classified \cite{Cappelli:1986hf,Cappelli:1987xt,Kato:1987td}. They are labeled by ADE Dynkin diagram $G$ with dual Coxeter number $g=k+2$. The modular invariant partition function is expressed using $N^{G}_{r,\bar{r}}$ as
\begin{align}
Z^G = \sum_{r , \bar{r}}N^{G}_{r,\bar{r}} \chi_{r} \chi_{\bar{r}},
\end{align}
where $\chi_{r}$ are the characters of affine Lie algebra $\widehat{su}(2)_k$.

The boundary conditions in su(2) WZW theory which preserve $\widehat{su}(2)_k$ are also all classified\cite{Behrend:1999bn}. Ishibashi states $|r\rangle\rangle$ are labeled by the finite set
$\mathcal{E}= \{ r : \Hcal_r \otimes \overline{\Hcal_r} \in \Hcal \} $. Boundary states $|a\rangle$  in the theory are
\begin{align}
|a\rangle  = \sum_{r \in \mathcal{E}} \frac{\psi^r_a(G)}{\sqrt{S^{(k)}_{1r}}} |r\rangle\rangle,
\end{align}
where $a$ is the label of the Dynkin diagram nodes.
The modular transformation S-matrix is $S^{(k)}_{ij}= \sqrt{\frac{2}{k+2}} \sin \frac{i j \pi}{k+2}$. ``Intertwiners'' $n^{r}_{a,b}(G)$ are defined as follow.
\begin{align}
Z_{a|b} = \langle a | \tilde{q}^{\frac{1}{2}(L_0 + \bar{L}_0 - \frac{c}{12})}|b\rangle = \sum_{r} n^{r}_{a,b}(G) \chi_{r} (q),
\end{align}
where $n^{r}_{a,b}(G)$ are obtained as
\begin{align}
n^{r}_{a,b}(G) = \sum_{r' \in \mathcal{E}} \frac{(\psi_{a}^{r'}(G))^* \psi_{b}^{r'}(G)}{S^{(k)}_{1r'}}S^{(k)}_{r'r} .
\label{Gspectrum}
\end{align}
$n^r_{a,b}(G)$ are non-negative integers. Note that $\psi_{a}^{r}$ satisfies the relation
\begin{align}
\psi_{a}^{r}(G) = (-1)^{\tau(a)} \psi_{a}^{g-r}(G),
\end{align}
for appropriately chosen $\tau(a)$.

$A,D_{\text{odd}},E_6$ has graph automorphism $\gamma$ and it satisfies the relation
\begin{align}
 \psi^{r}_{\gamma(a)}(G)=(-1)^{r+1}\psi^{r}_{a}(G).
\end{align}

Here we summarize these quantities for $D_6$ and $E_6$ diagram. In the Dynkin diagram, we express the value of $\tau$ by the colored nodes as $\tau(\circ)=0,\tau(\bullet)=1$.
\begin{itemize}
 \item $D_6$ : $g=10$ \qquad
 \begin{minipage}{5cm}
\includegraphics[scale=0.8]{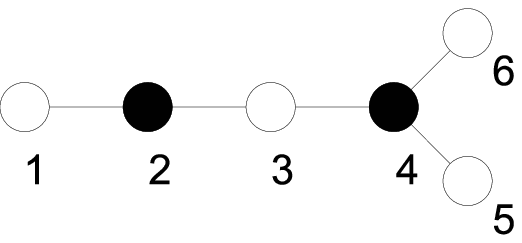}
 \end{minipage}
  \begin{itemize}
    \item Modular invariant
$ |\chi_1 + \chi_9|^2 + |\chi_3 + \chi_7 |^2 + 2 |\chi_5 |^2$.
   \item Exponents $\mathcal{E}(D_6)=\{1,3,5,5',7,9\}$.
   \item Boundary state coefficients $\frac{\psi^r_a(D_6)}{\sqrt{S^{(8)}_{1r}}}$
\begin{align}
 \begin{array}{c|cccccc}
 a\backslash r &1 &3 &5 &5' &7 &9 \\ \hline
1 & \sqrt{2\alpha_{-}}&\sqrt{2\alpha_{+}} &5^{-1/4} &5^{-1/4}
                                         &\sqrt{2\alpha_{+}} &\sqrt{2\alpha_{-}} \\
2 & 1&1 &0 &0 &-1 &-1 \\
3&  \sqrt{40\alpha_{+}^3}&\sqrt{40\alpha_{-}^3} &-5^{-1/4} &-5^{-1/4} &\sqrt{40\alpha_{-}^3} &\sqrt{40\alpha_{+}^3} \\
 4& \sqrt{20}\alpha_{+}&-\sqrt{20}\alpha_{-} & 0 & 0 &\sqrt{20}\alpha_{-} &-\sqrt{20}\alpha_{+} \\
 5& \sqrt{2\alpha_{+}}&-\sqrt{2\alpha_{-}} &2\cdot5^{1/4} \alpha_{+}&-2\cdot5^{1/4} \alpha_{-}
                                         &-\sqrt{2\alpha_{-}} &\sqrt{2\alpha_{+}} \\
 6& \sqrt{2\alpha_{+}}&-\sqrt{2\alpha_{-}} &-2\cdot5^{1/4} \alpha_{-}&2\cdot5^{1/4} \alpha_{+}
                                         &-\sqrt{2\alpha_{-}} &\sqrt{2\alpha_{+}} \\
 \end{array}
\end{align}
where $\alpha_{\pm} := \frac{5\pm \sqrt{5}}{20}$.
  \end{itemize}
 \item $E_6$ : $g=12$ \qquad
 \begin{minipage}{5cm}
\includegraphics[scale=0.8]{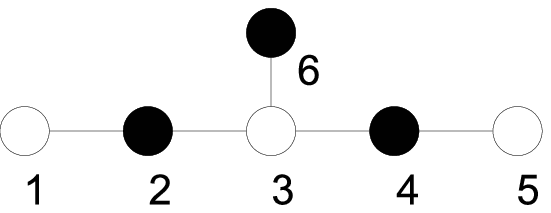}
 \end{minipage}
   \begin{itemize}
    \item Modular invariant
       $|\chi_1 + \chi_7|^2 + |\chi_4 + \chi_8 |^2 +  |\chi_5 + \chi_{11}|^2$.
    \item Exponents $\mathcal{E}(E_6)=\{1,4,5,7,8,11\}$.
    \item $\gamma: (1,2,3,4,5,6)\to (5,4,3,2,1,6)$.
    \item Boundary state coefficients $\frac{\psi^s_b(E_6)}{\sqrt{S^{(10)}_{1s}}}$
\begin{align}
  \begin{array}{c|cccccc}
  b\backslash s &1 &4 &5 &7 &8 &11 \\ \hline
1&   \frac{1}{\sqrt 2} & 2^{-1/4} &\frac{1}{\sqrt 2} &\frac{1}{\sqrt 2} &2^{-1/4} &\frac{1}{\sqrt 2} \\
 2&  4\sqrt3 \beta_{+}^2&2^{-1/4} &4\sqrt3 \beta_{-}^2
                    &-4\sqrt3 \beta_{-}^2 &-2^{-1/4} & -4\sqrt3 \beta_{+}^2 \\
 3&  4\sqrt6 \beta_{+}^2& 0 &-4\sqrt6 \beta_{-}^2
                    &-4\sqrt6 \beta_{-}^2 & 0 & 4\sqrt6 \beta_{+}^2 \\
  4& 4\sqrt3 \beta_{+}^2& -2^{-1/4} & 4\sqrt3 \beta_{-}^2
                    &-4\sqrt3 \beta_{-}^2 & 2^{-1/4} & -4\sqrt3 \beta_{+}^2 \\
 5&  \frac{1}{\sqrt 2} & -2^{-1/4} &\frac{1}{\sqrt 2} &\frac{1}{\sqrt 2}& -2^{-1/4} &\frac{1}{\sqrt 2} \\
 6&  1&0 &-1 &1 &0 &-1 \\
  \end{array}
\end{align}
where $\beta_{\pm} :=\frac{1}{2}\sqrt{\frac{3\pm\sqrt{3}}{6}}$.
   \end{itemize}
\end{itemize}

The following relations between intertwiners are useful. The $D_6$ intertwiners $n^{r}_{a,a'}(D_6)$ satisfy\footnote{Similar relations for $D_{\text{even}}, E_7,E_8$ are also satisfied.}
\begin{align}
 n^{r}_{a,a'}(D_6)=n^{10-r}_{a,a'}(D_6).\label{n-relation0}
\end{align}
It is convenient to use the function $\rho:\{3,6\}\to\{1,2\}$ as
\begin{align}
 \rho(3):=2,\qquad \rho(6):=1.\label{def-rho}
\end{align}
The $E_6$ intertwiners satisfy for $b,b'\in \{3,6\}$
\begin{align}
&n^{s}_{b,b'}(E_6)=n^{s}_{\rho(b),\rho(b')}(E_6)+n^{12-s}_{\rho(b),\rho(b')}(E_6),\\
&n^{s}_{b,\rho(b')}(E_6)=n^{s}_{\rho(b),b'}(E_6),
\end{align}
and for $b=3,6$ and arbitrary $b'$
\begin{align}
 n^{s}_{b,b'}(E_6)=n^{12-s}_{b,b'}(E_6).\label{n-relation1}
\end{align}
\subsection{Boundary states of $D_6-E_6$ theory}
There are 36 Ishibashi sates in the $D_6 - E_6$ theory (3.2) which satisfy the superconformal gluing conditions.
\begin{align}
T(z) = \bar{T}(\bar{z}) , \quad G(z) = \eta \bar{G}(\bar{z}).
\end{align}
Ishibashi states for $(r,t,s)$ module are denoted by $|(r,t,s)_{10}\rangle\rangle$. Since there are two degeneracy ( $2$ and $\tilde{2}$ ) in $R$-sector, the indices $r,t,s$ take values in
\begin{align}
r \in \{ 1,3,5,5',7,9\} ,\qquad
t \in \{ 1,2,\tilde{2},3 \} , \qquad
s \in \{ 1,4,5,7,8,11 \}.
\end{align}
Note that
\begin{align}
| (r,t,s)_{10} \rangle \rangle = |(10-r,4-t,12-s)_{10}\rangle\rangle.
\end{align}
Since the exponents of $D_6$ are all odd number, $s$ is odd in NS-sector and $s$ is even in R-sector. Taking the above identification into account, $s=1,5$ in NS-sector and $s=4$ in R-sector.

These Ishibashi states satisfy the relation
\begin{align}
\langle \langle  (r,t,s)_{10} |\tilde{q}^{{1 \over 2} (L_0 + \bar{L}_0-{c\over 12})}| (r',t',s')_{10}\rangle \rangle &=  \delta_{[r,t,s],[r',t',s']} \chi^{(10)}_{r,t,s}(\tilde{q}).
\end{align}
$[r,t,s]$ represents an equivalence class under the relation $\sim$ in \eqref{equiv}.

Modular S transformation rule of the character $\chi^{(10)}_{r,t,s}$ is given by
\begin{align}
\chi^{(10)}_{r,t,s} (\tilde{q})= \sum_{r'=1}^{9} \sum_{t'=1}^{3}\sum_{s'=1}^{11} S^{(8)}_{r'r} S^{(2)}_{t't} S^{(10)}_{s's} \chi^{(10)}_{r',t',s'} (q) =\sum_{[r',t',s']} 2  S^{(8)}_{r'r} S^{(2)}_{t't} S^{(10)}_{s's}  \chi^{(10)}_{r',t',s'} (q) .
\end{align}
where $S^{(k)}_{r'r}$ is the S-matrix of $\widehat{SU}(2)_k$.

Let us introduce the following notation for the states.
\begin{align}
|a,b;NS \rangle &= \sum_{r\in {\cal E}(D_6),\ s=1,5} \frac{\psi_a ^{r}(D_6) \psi_b^{s}(E_6)}{\sqrt{S^{(8)}_{1r} S^{(10)}_{1s}}} (|(r,1,s)_{10}\rangle\rangle + |(r,3,s)_{10}\rangle\rangle), \nonumber
\\
|a,b;\widetilde{NS} \rangle &= \sum_{r\in {\cal E}(D_6),\ s=1,5} \frac{\psi_a ^{r}(D_6) \psi_b^{s}(E_6)}  {\sqrt{S^{(8)}_{1r} S^{(10)}_{1s}}} (|(r,1,s)_{10}\rangle\rangle - |(r,3,s)_{10}\rangle\rangle), \nonumber
\\
|a,b;R\rangle &= \sum_{r\in {\cal E}(D_6),\ s=4} 2^{1/4} \frac{\psi_a ^{r}(D_6) \psi_b^{s}(E_6)} {\sqrt{S^{(8)}_{1r} S^{(10)}_{1s}}}  |(r,2,s)_{10}\rangle \rangle,  \nonumber
\\
|a,b;\widetilde{R} \rangle &= \sum_{r\in {\cal E}(D_6),\ s=4} 2^{1/4}\frac{\psi_a ^{r}(D_6) \psi_b^{s}(E_6)} {\sqrt{S^{(8)}_{1r} S^{(10)}_{1s}}}  |(r,\tilde{2},s)_{10}\rangle \rangle.
\label{Bstates0}
\end{align}
These states satisfy the relations
\begin{align}
 &|a,\gamma(b),NS\rangle=+|a,\gamma(b),NS\rangle,\qquad
 |a,\gamma(b),\widetilde{NS}\rangle=+|a,\gamma(b),\widetilde{NS}\rangle,\nonumber\\
 &|a,\gamma(b),R\rangle=-|a,\gamma(b),R\rangle,\qquad
 |a,\gamma(b),\widetilde{R}\rangle=-|a,\gamma(b),\widetilde{R}\rangle.
\end{align}
In particular
\begin{align}
 |a,3,R\rangle=
 |a,6,R\rangle=
 |a,3,\widetilde{R}\rangle=
 |a,6,\widetilde{R}\rangle=0.
\end{align}
We introduce the notation $n^{r,t,s}_{A,B}$ as open string spectrum between two states $|A\rangle$ and $|B\rangle$ as follows.
\begin{align}
\langle A | \tilde{q}^{{1 \over 2} (L_0 + \bar{L}_0-{c\over 12})} |B\rangle = \sum_{[r,t,s]} n^{r,t,s}_{A,B} \chi_{r,t,s}(q).
\end{align}
These $n$'s between two states in \eqref{Bstates0} can be calculated using the properties on $\psi$ \eqref{Gspectrum}.
\begin{align}
&n^{r,t,s}_{(a,b;NS),(a',b';NS)} = n^{r,t,s}_{(a,b;\widetilde{NS}),(a',b';\widetilde{NS})} = \left\{
      \begin{array}{ll}
        0, & \hbox{$t$=2}, \\
        \frac{1}{2}(n^r_{a,a'}(D_6)n^{s}_{b,b'}(E_6)+n^{10-r}_{a,a'}(D_6)n^{12-s}_{b,b'}(E_6)), & \hbox{$t$=1,3},
      \end{array}
    \right. \nonumber
\\
&n^{r,t,s}_{(a,b;R),(a',b';R)} = n^{r,t,s}_{(a,b;\widetilde{R}),(a',b';\widetilde{R})}
= \left\{
      \begin{array}{ll}
         \frac {1}{2}(n^r_{a,a'}(D_6)n^{s}_{b,b'}(E_6)-n^{10-r}_{a,a'}(D_6)n^{12-s}_{b,b'}(E_6)), & \hbox{$t$=1}, \\
         -\frac{1}{2}(n^r_{a,a'}(D_6)n^{s}_{b,b'}(E_6)-n^{10-r}_{a,a'}(D_6)n^{12-s}_{b,b'}(E_6)), & \hbox{$t$=3}, \\
        0, & \hbox{$t$=2},
      \end{array}
    \right. \nonumber
\\
&n^{r,t,s}_{(a,b;NS),(a',b';\widetilde{NS})}
= \left\{
      \begin{array}{ll}
        \frac{\sqrt{2}}{2}(n^r_{a,a'}(D_6)n^{s}_{b,b'}(E_6)+n^{10-r}_{a,a'}(D_6)n^{12-s}_{b,b'}(E_6)), & \hbox{$t$=2}, \\
        0, & \hbox{$t$=1,3},
      \end{array}
    \right. \nonumber
\\
&n^{r,t,s}_{(a,b;R),(a',b';\widetilde{R})}
=0.
\label{overlap1}
\end{align}

For this $D_6 - E_6$ theory, we find two consistent sets of 36 boundary states. The first set consists of the following 36 boundary states.
\begin{align}
\textrm{type I } &: |a,b;NS\rangle + |a,b;R\rangle,
\qquad (a=1,3,5,6,\ b=1,3,5) \text{ or } (a=2,4,\ b=2,4,6), \nonumber
\\
\textrm{type II } &: \sqrt{2}|a,b;\widetilde{NS} \rangle,
\qquad(a,b)=(1,2),(3,2),(5,2),(6,2),(2,1),(4,1),  \nonumber
\\
\textrm{type III } &: \frac{1}{\sqrt{2}} |a,b;\widetilde{NS}\rangle + |a,\rho (b); \widetilde{R}\rangle, \qquad (a,b)=(1,6),(3,6),(5,6),(6,6),(2,3),(4,3),  \nonumber
\\
\textrm{type IV } &: \frac{1}{\sqrt{2}} |a,b;\widetilde{NS}\rangle - |a,\rho (b); \widetilde{R}\rangle, \qquad (a,b)=(1,6),(3,6),(5,6),(6,6),(2,3),(4,3).
\label{36boundary1}
\end{align}
Here, $\rho$ is defined in eq.\eqref{def-rho}. The other set consists of
\begin{align}
\textrm{type i } &: |a,b;NS\rangle + |a,b;R\rangle, \qquad (a=1,3,5,6,\ b=2,4,6) \text{ or } (a=2,4,\ b=1,3,5),\nonumber
\\
\textrm{type ii } &: \sqrt{2}|a,b;\widetilde{NS} \rangle,  \qquad (a,b)=(1,1),(3,1),(5,1),(6,1),(2,2),(4,2),
\nonumber
\\
\textrm{type iii } &: \frac{1}{\sqrt{2}} |a,b;\widetilde{NS}\rangle + |a,\rho (b); \widetilde{R})\rangle, \qquad (a,b)=(1,3),(3,3),(5,3),(6,3),(2,6),(4,6),
\nonumber
\\
\textrm{type iv } &: \frac{1}{\sqrt{2}} |a,b;\widetilde{NS}\rangle - |a,\rho (b); \widetilde{R})\rangle, \qquad(a,b)=(1,3),(3,3),(5,3),(6,3),(2,6),(4,6).
\label{36boundary2}
\end{align}
The Cardy condition can be checked by using \eqref{overlap1} and the relations \eqref{n-relation0}--\eqref{n-relation1}. For the first set of boundary states \eqref{36boundary1}, non-zero $n$'s are
\begin{align}
n^{r,1,s}_{I,I} &= n^r_{a,a'} (D_6) n^s_{b,b'}(E_6), \nonumber
\\
n^{r,2,s}_{I,II} &= n^r_{a,a'}(D_6) n^s_{b,b'}(E_6) + n^r_{a,a'}(D_6) n^{12-s}_{b,b'}(E_6), \nonumber
\\
n^{r,2,s}_{I, III} = n^{r,2,s}_{I,IV} &= n^r_{a,a'}(D_6) n^s_{b,b'} (E_6), \nonumber
\\
n^{r,1,s}_{II,II} &= n^r_{a,a'} ( D_6 ) n^s_{\rho^{-1}(b),\rho^{-1} (b')} (E_6), \nonumber
\\
n^{r,1,s}_{II,III} = n^{r,1,s}_{II,IV} &= n^r_{a,a'} (D_6) n^s_{b,b'}(E_6) , \nonumber
\\
n^{r,1,s}_{III,III} = n^{r,1,s}_{IV,IV} &= n^r_{a,a'} (D_6) n^s_{\rho(b), \rho(b')} (E_6) , \nonumber
\\
n^{r,1,s}_{III,IV} &= n^r_{a,a'} (D_6) n^{12-s}_{\rho(b), \rho(b')} (E_6),
\nonumber\\
n^{r,3,s}_{X,Y}&=n^{10-r,1,12-s}_{X,Y}.
\label{n_36_boundary}
\end{align}
\eqref{n_36_boundary} is still valid for the second set of boundary states \eqref{36boundary2}.


\providecommand{\href}[2]{#2}\begingroup\raggedright\endgroup

\end{document}